\documentclass[oneside, a4paper, onecolumn, 11pt]{article}

\usepackage[left=2cm,top=2cm,bottom=1.5cm,right=2cm]{geometry}
\usepackage{hyperref}
\usepackage[utf8]{inputenc}
\usepackage{graphicx} 		
\usepackage{amsmath}  		
\usepackage{natbib}	
\usepackage[official]{eurosym}
\usepackage{tabularx}
\usepackage{booktabs}
\usepackage{parskip}
\usepackage{tikz}
\usepackage[verbose]{wrapfig}
\usetikzlibrary{shapes,arrows,positioning,fit,backgrounds,calc}
\usepackage{bclogo}
\usepackage{environ}
\usepackage{lipsum}
\usepackage{enumitem}
\setitemize{noitemsep,topsep=0pt,parsep=0pt,partopsep=0pt,leftmargin=*}
\usepackage{amssymb}

\usepackage{nopageno}
\usepackage{enumitem}

\usepackage{tabularx}
\usepackage{booktabs}
\usepackage{longtable}

\usepackage{fancyhdr}
\newenvironment{itemize*}%
  {\begin{itemize}%
    \setlength{\itemsep}{0pt}%
    \setlength{\parskip}{0pt}}%
  {\end{itemize}}

\NewEnviron{mybox}[1]
  {\wrapfigure{r}{0.5\textwidth}
  \setlength\fboxrule{1.pt}
  \fcolorbox{gray!60}{white}{%
  \begin{minipage}{\dimexpr\linewidth-2\fboxsep-2\fboxrule\relax}
  \parbox[t][0.5cm][t]{\dimexpr\linewidth-0.5cm\relax}{\bfseries#1}
  \BODY
  \end{minipage}}%
\endwrapfigure}

\usepackage{enumitem}

\author{Di Molfetta\\ Quantum Technology Consultant}

\begin{document}

\

\section*{Quantum Computing : A New Frontier for Science and Society}

The dawn of the 21st century has brought with it a technological revolution that promises to redefine the boundaries of computation, communication, and scientific discovery: \textit{quantum technologies}. Rooted in the principles of quantum mechanics these technologies are poised to unlock solutions to problems that have long eluded classical approaches. For instance, quantum computers could simulate molecular interactions with unprecedented accuracy, potentially revolutionizing drug discovery and materials science~\citep{Feynman1982,preskill2018nisq}. Beyond computation, \textit{quantum sensors} and \textit{communication networks} are already demonstrating capabilities far beyond classical limits, such as ultra-precise measurements and theoretically unhackable data transmission~\citep{Dowling2003,Gisin2007}.

But what are the physical resources behind such a revolution? Quantum computing derives their computational advantages from three fundamental quantum resources that have no classical analogues~: superposition, entanglement, and interference. These resources enable quantum algorithms to achieve exponential or polynomial speedups over their classical counterparts for specific problem classes where quantum effects can be effectively harnessed \citep{nielsen2010quantum}. The most fundamental quantum resource is \textit{superposition}, which allows quantum bits to exist in combinations of multiple states simultaneously. While classical bits are strictly binary ($0$ or $1$), qubits can represent complex combinations of basis states, enabling \textit{parallel exploration of computational paths}. This quantum parallelism constitutes the key for evaluating multiple potential solutions. Quantum interference, at this point helps to select and to amplify the correct one, such as in the Grover algorithm~\citep{grover1996fast}. This comes from the fact that in quantum computing we deal with complex amplitudes (vectors), differently from the classical framework where we manipulate only non-negative numbers (probabilities, $p>0$). This would mean that, all along our quantum calculation, one may arrive to engineer relative phases so that probability amplitudes for wrong answers cancel (destructive interference), while amplitudes for right answers add up (constructive interference). The speedup is not “parallelism” by itself (a superposition does not let you 'read' all branches), but the ability to shape the final measurement distribution by coherently combining many computational paths.
This intrinsic wave-behaviour of quantum particles is indeed the fundamental reason for quantum computational advantage in many family of quantum algorithms, allowing to \textit{search} solution spaces more efficiently than classical approaches for certain \textit{difficult} problems. Obviously interference gives an advantage only if the computation preserves and controls (complex) phases. This means that coherence time must exceed circuit depth (otherwise phases randomize and  the interference disappears), the gates must have sufficient fidelity to maintain phase relations,and finally the algorithm must have a structure where the desired answer is a global property (period, marked set, eigenphase, etc.) that can be amplified by coherent summation.

Entanglement represents another crucial quantum resource that creates non-classical and \textit{non-local} correlations between distinct qubits. It arises when the quantum state of a composite system cannot be factorized into independent states of its subsystems. In such states, measurements performed on one subsystem influence the statistical outcomes of measurements on another, even when the subsystems are spatially separated. These correlations violate classical bounds such as \textit{Bell inequalities} and constitute a key resource for quantum information processing \cite{Einstein1935,Bell1964,nielsen2010quantum}. Entanglement enables several protocols with no classical analogue, including quantum teleportation, in which the quantum state of a particle can be transferred between distant locations using a shared entangled state and classical communication \cite{Bennett1993}. Such states also serve as the foundation for secured quantum communication protocols like \textit{quantum key distribution}~\citep{ekert1991quantum}, and for distributed quantum computation where multiple quantum processors, sharing a non local quantum state, can work in concert~\citep{horodecki2009quantum}, leading to message and communication advantage with respect to 
classical routing. 

However, realizing these advantages faces significant practical challenges. Quantum states are extremely fragile, especially with respect to thermal noise and more in general to any possible physical interaction (even with cosmic ray particles). Quantum systems may lose their quantum properties in a characteristic time, which we call \textit{decoherence time}. This fundamentally limits the depth of quantum circuits that can be executed before errors accumulate~\citep{preskill1998reliable}. Unfortunately current error correction schemes require substantial physical qubit overhead to implement each logical qubit, with estimates suggesting that thousands of physical qubits may be needed for each fault-tolerant logical qubit~\citep{fowler2012surface}. The interface between classical and quantum systems also presents bottlenecks, as quantum information must be carefully prepared and measured, limiting the rate at which problems can be loaded and solutions read out.

Moreover the computational advantages of quantum computing are problem-specific and require careful algorithm design to exploit quantum resources effectively~\citep{huang2025vast}. While not universally faster than classical computing, quantum algorithms provide measurable advantages for carefully selected problems where the unique properties of superposition, entanglement, and interference can be harnessed. As hardware improves and error correction techniques mature, these quantum advantages will become increasingly practical for applications in cryptography, optimization, simulation, and other domains where classical algorithms face fundamental limitations~\citep{eisert2025mind}.

This report explores the current state of quantum technologies, their potential applications, and the challenges that must be addressed to harness their full potential. In particular we will focus on the \textit{quantum computer architecture and its ecosystem}. Such architecture represents a complex, multi-layered system that integrates quantum and classical components to enable the execution of quantum algorithms. Our manuscript will be then organised as follows~: first we will introduce the quantum processing unit, the lowest layer of a quantum computer. Then we will progress from the lowest to the higher layer of the system architectures : measurement, circuit control, error correction and mitigation system, the quantum compiler and finally the software stack, with particular emphasis on the interactions between these components. 

\begin{figure}[t]
\centering
\includegraphics[width=\linewidth]{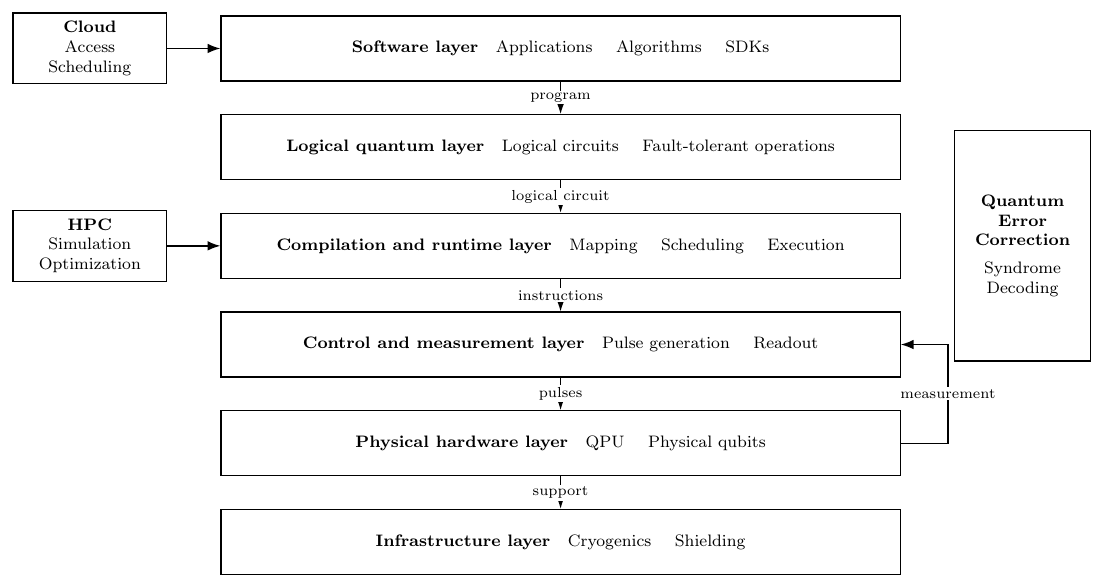}
\caption{Layered architecture of a quantum computing system showing the interaction between the software stack, compilation and runtime layers, control electronics, and the quantum processor operating within a cryogenic infrastructure. Quantum error correction spans several layers, while HPC and cloud resources provide classical support.}
\label{fig:quantum_architecture}
\end{figure}

\subsection*{1 Quantum Processing Unit Architecture}

The quantum processing unit (QPU) represents the key part of a quantum computer, the quantum hardware, where quantum information is stored and processed. The QPU consists of an array of \textit{physical qubits} implemented using various physical systems, each presenting unique advantages and challenges in terms of coherence times, gate fidelities, and connectivity. 
Over the past decade, researchers and companies have explored a variety of quantum hardware platforms. Here we delve into the leading qubit technologies---superconducting qubits, trapped ions, neutral atoms, photonic qubits, topological qubits, and silicon spin qubits---examining their advantages, disadvantages, and the most recent milestones that are shaping the future of quantum computing.

\subsubsection*{Superconducting Qubits}

Superconducting qubits have emerged as the leading platform for near-term quantum computation, leveraging the macroscopic quantum coherence of superconducting circuits operated at cryogenic temperatures. These qubits, typically fabricated from aluminum or niobium and cooled to 10-20 mK, encode quantum information in the phase degree of freedom of superconducting circuits, with the transmon qubit being the most common implementation~\citep{kjaergaard2020superconducting}. The transmon design provides robust protection against charge noise while maintaining sufficient anharmonicity for selective control of the qubit subspace~\citep{schreier2008suppressing}. Control is achieved through microwave pulses resonant with the qubit's transition frequency, with coupling between qubits implemented via capacitive or inductive interactions or through microwave resonators \citep{barends2014superconducting}.

The primary advantages of superconducting qubits stem from their compatibility with established semiconductor fabrication techniques, enabling rapid iteration and scalability. This has facilitated the development of large qubit arrays, with IBM's 1,121-qubit Condor processor representing the current state of the art in qubit count~\citep{jurcevic2023ibm}. Superconducting qubits also benefit from fast gate operations on nanosecond timescales and mature control electronics infrastructure. Recent advancements have pushed single-qubit gate fidelities above 99.99\% and two-qubit gate fidelities beyond 99.9\%, approaching the thresholds required for fault-tolerant quantum computation \citep{place2021superconducting}.

However, superconducting qubits face significant challenges related to their sensitivity to thermal noise and material defects, necessitating operation in complex cryogenic environments. Decoherence remains a critical limitation, though recent improvements in materials engineering and qubit designs have extended coherence times. The development of dynamic circuit execution has enabled real-time error mitigation during computation, representing a significant step toward practical quantum algorithms \citep{jurcevic2023ibm}. Google's demonstration of a logical qubit with error rates below the fault-tolerance threshold using surface codes marks another major milestone in the field \citep{ai2023exponential}.

\subsubsection*{Trapped Ion Qubits}

Trapped ion qubits represent one of the most mature quantum computing platforms, utilizing individual atomic ions confined in electromagnetic traps. Typically employing ions such as \(^{171}\text{Yb}^+\) or \(^{43}\text{Ca}^+\), these systems encode quantum information in hyperfine or optical electronic states that exhibit exceptional coherence properties~\citep{bruzewicz2019trapped}. The ions are confined using Paul traps or Penning traps, where oscillating electric fields or static magnetic/electric fields create harmonic potential wells for precise spatial manipulation. Quantum gates are implemented using laser-induced Raman transitions for single-qubit operations and Mølmer-Sørensen or Cirac-Zoller gates for two-qubit entanglement~\citep{sorensen1999entanglement}.

The trapped ion platform offers several distinctive advantages, including exceptionally long coherence times reaching seconds, high-fidelity quantum operations exceeding 99.99\% for single-qubit gates, and all-to-all connectivity via shared motional modes. These properties make trapped ions particularly suitable for applications requiring precise quantum state control. Recent breakthroughs have demonstrated logical qubits with error rates below \(10^{-6}\), achieved through surface code error correction across multiple physical qubits \citep{hild2022fault}. IonQ's 256-qubit Aria system has set new records with quantum volume exceeding 4 million, showcasing the platform's potential for large-scale quantum computation~\citep{ionq2023aria}.

Despite these strengths, trapped ion systems face challenges related to slow gate operations on microsecond timescales and the complexity of laser control requirements. Scalability remains an ongoing challenge, though modular architectures with photonic interconnects between separate ion traps have shown promise for overcoming single-trap limitations \citep{monroe2023modular}. Recent demonstrations of quantum advantage for specific optimization problems highlight the platform's potential for near-term practical applications~\citep{ebadi2023quantum}.

\subsubsection*{Neutral Atom Qubits}

Neutral atom qubits have emerged as a rapidly advancing platform for quantum computation, utilizing arrays of individual atoms trapped in optical potentials. These systems typically employ alkali metals such as rubidium or cesium, confined in optical tweezers or optical lattices, where quantum information is encoded in hyperfine or optical ground states~\citep{ebadi2022quantum}. Single-qubit operations are performed via Raman transitions, while two-qubit gates are implemented through Rydberg blockade interactions, where atoms in highly excited Rydberg states prevent nearby atoms from being excited~\citep{saffman2010quantum}.

The neutral atom platform offers several compelling advantages, including long coherence times typically exceeding seconds, scalable architectures enabled by optical tweezer arrays, and parallel gate operations through global laser addressing. Recent advancements have demonstrated programmable 256-qubit processors with high-fidelity operations, showcasing the platform's potential for large-scale quantum computation~\citep{ebadi2023quantum}. The system has also demonstrated quantum advantage for specific optimization problems using 200+ qubit systems, leveraging parallel gate operations and reconfigurable connectivity~\citep{henriet2024quantum}.

Current research focuses on improving error correction protocols, developing hybrid architectures with photonic interconnects, and implementing mid-circuit measurement capabilities. Recent breakthroughs in error mitigation techniques, including dynamical decoupling, have extended coherence times and improved computational reliability~\citep{bluvstein2024error}. The development of mid-circuit measurement and feedback control enables adaptive quantum algorithms and enhances error correction capabilities~\citep{saffman2024midcircuit}.

\subsubsection*{Photonic Qubits}

Photonic qubits encode quantum information in the properties of light, typically using polarization, phase, or time-bin encoding, offering unique advantages for quantum communication and distributed quantum computing. These qubits are generated using spontaneous parametric down-conversion or quantum dots, and manipulated using linear optical elements and detectors~\citep{obrien2009photonic}. Photonic systems operate at room temperature and are particularly suited for quantum communication applications due to their compatibility with existing fiber-optic infrastructure.

The primary advantages of photonic qubits include natural resilience to decoherence at room temperature, high-speed operations at terahertz frequencies, and long-distance transmission capabilities. Recent breakthroughs have demonstrated quantum advantage in Gaussian boson sampling using Xanadu's 216-qubit Borealis processor, performing computations believed to be intractable for classical supercomputers~\citep{xanadu2023borealis}. The platform has also seen progress in developing million-qubit photonic quantum computers using silicon photonics, aiming to leverage existing semiconductor manufacturing infrastructure for scalability \citep{psiquantum2023}.

However, photonic qubits face significant challenges in implementing deterministic two-qubit gates, which are essential for universal quantum computation. While linear optical quantum computing can achieve probabilistic gates using ancillary photons and post-selection, this approach remains resource-intensive and limits scalability. Current research focuses on developing photonic integrated circuits to address these challenges and improve the platform's practicality for quantum computation~\citep{wang2023photonic}.

\subsubsection*{Topological Qubits}

Topological qubits represent a theoretically robust approach to quantum computation, encoding quantum information in the non-local properties of quasiparticles such as Majorana fermions or anyons. These qubits leverage topological protection, where quantum information is encoded in the global properties of the system rather than local degrees of freedom, providing inherent robustness against local noise~\citep{nayak2008non}. The most promising implementations involve superconducting circuits or semiconductor nanowires with strong spin-orbit coupling, though experimental realization remains challenging.

The primary advantage of topological qubits lies in their inherent robustness against local noise, which could dramatically reduce the overhead required for error correction. This makes them a leading candidate for fault-tolerant quantum computation, as they could potentially achieve ultra-low error rates without extensive error correction. Recent progress has provided evidence of non-Abelian anyons in two-dimensional electron gases, representing a significant step toward experimental verification of topological quantum computation~\citep{bartolomei2023anyons}.

However, topological qubits face significant experimental challenges, including the need for extremely pure materials and precise control at the nanoscale. The operational speed of topological qubits is expected to be slower than other platforms due to the requirement for physical braiding operations. While no company or research group has conclusively demonstrated a functional topological qubit, ongoing research explores hybrid architectures that combine topological protection with other qubit technologies~\citep{groszkowski2024topological}.

\subsubsection*{Silicon Spin Qubits}

Silicon spin qubits encode quantum information in the spin states of electrons or nuclei within silicon-based devices, leveraging the mature semiconductor manufacturing infrastructure. Quantum information is typically encoded in the spin states of electrons in quantum dots or donors in silicon, with control achieved using microwave pulses and electrostatic gates, and readout performed via spin-dependent tunneling or charge sensing~\citep{vandersypen2017interfacing}.

The silicon spin qubit platform offers several advantages, including compatibility with existing semiconductor fabrication techniques, higher operating temperatures (1-4 K) compared to superconducting qubits, and long coherence times due to isotopic purification. Recent advancements have demonstrated high-fidelity single- and two-qubit gates exceeding 99.9\%, with Intel's 12-qubit Tunnel Falls chip representing a significant milestone in the field \citep{intel2023tunnel}. The platform has also shown progress in coherent control of multi-qubit processors, with demonstrations of 6-qubit systems showcasing potential for scalability \citep{xue2022silicon}.

However, silicon spin qubits face challenges related to limited qubit connectivity and slower gate operations compared to other platforms. The need for precise control of large arrays adds complexity to the system. Current research focuses on improving qubit connectivity through shuttle-based architectures and enhancing gate fidelities to match those of leading platforms~\citep{zwerver2022silicon}.

\subsubsection*{Comparative Analysis and Future Outlook}

The landscape of quantum hardware is diverse, with each platform offering unique trade-offs between \textbf{coherence time, gate fidelity, scalability, and operational complexity}. Superconducting qubits lead in scalability and near-term applicability, but they struggle with error rates and cooling requirements. Trapped ions and neutral atoms excel in precision and coherence, but their slow gate operations and scalability challenges remain hurdles. Photonic qubits offer room-temperature operation and natural resilience to decoherence, but they face difficulties in implementing deterministic gate. Topological qubits promise unprecedented fault tolerance, but their experimental realization is still uncertain. Silicon spin qubits leverage existing semiconductor infrastructure, but they require advances in connectivity and gate fidelity. 

Looking ahead, the future of quantum hardware will likely involve hybrid architectures that combine the strengths of multiple platforms. For example, superconducting or silicon qubits could be used for local processing, while photonic interconnects enable long-distance communication between quantum processors. Meanwhile, advances in error correction, materials science, and control electronics will be critical to overcoming the limitations of current systems.

\begin{table}[h!]
\centering
\scriptsize
\caption{Comparison of Major Quantum Hardware Platforms}
\label{tab:quantum_hardware}
\renewcommand{\arraystretch}{1.3}
\begin{tabularx}{\textwidth}{l X X X X}
\toprule
\textbf{Hardware Platform} & \textbf{Physical Principle} & \textbf{Advantages} & \textbf{Disadvantages} & \textbf{Recent Results (2023--2025)} \\
\midrule
Superconducting qubits &
Josephson junction circuits operating at millikelvin temperatures &
Fast gate operations; mature fabrication methods; strong industry support (IBM, Google) &
Shorter coherence times; requires cryogenic infrastructure; wiring and scaling complexity &
Processors exceeding 1000 qubits (IBM Condor); demonstrations of below-threshold error correction on $\sim$100 qubits \\
\midrule
Trapped-ion qubits &
Individual ions confined in electromagnetic traps and controlled with lasers &
Very long coherence times; high gate fidelities ($>$99.9\%); all-to-all connectivity &
Slow gate speeds; complex laser systems; scaling is technically difficult &
Demonstrations of logical qubits with record fidelities; systems with 50+ ions commercially available \\
\midrule
Photonic qubits &
Quantum states encoded in single photons propagating in optical circuits &
Operate at room temperature; ideal for networking; low decoherence &
Probabilistic entanglement; difficult deterministic two-qubit gates; optical complexity &
Large-scale photonic processors with integrated silicon photonics; advances in cluster-state quantum computing \\
\midrule
Neutral-atom qubits &
Atoms trapped in optical tweezers and excited to Rydberg states &
Highly uniform qubits; flexible connectivity; scalable 2D and 3D arrays &
Complex optical control; sensitivity to laser noise; error rates still significant &
Demonstrations of programmable arrays exceeding 6000 atoms; improved two-qubit gate fidelities \\
\midrule
Spin qubits &
Electron or nuclear spins in quantum dots or defects in silicon &
Compatibility with CMOS fabrication; high density integration; small device footprint &
Sensitive to charge noise; readout and control remain challenging &
Multi-qubit silicon spin processors demonstrated; improved coherence using isotopically purified silicon \\
\midrule
Topological qubits &
Non-Abelian quasiparticles (Majorana zero modes) for intrinsically protected states &
Potential intrinsic fault tolerance; reduced error correction overhead &
Experimental realization not yet conclusive; extremely challenging fabrication &
First prototype devices investigating Majorana modes; no large-scale qubit systems yet demonstrated \\
\bottomrule
\end{tabularx}
\end{table}

\subsubsection*{Quantum metrics : how to evaluate a quantum hardware}

The performance of quantum computing hardware is determined by a complex interplay of physical coherence properties, operational accuracy, and architectural scalability. Unlike classical processors, quantum systems must preserve fragile quantum states while executing sequences of high-fidelity operations under strict noise constraints. Consequently, the characterization of quantum processors requires a diverse set of metrics that quantify both the intrinsic properties of individual qubits and the collective performance of multi-qubit systems.

At the device level, metrics such as the energy relaxation time ($T_1$), the phase coherence time ($T_2$), and noise spectral properties describe the susceptibility of qubits to decoherence mechanisms arising from interactions with the surrounding environment. At the operational level, gate fidelities, measurement fidelities, and leakage or crosstalk rates provide quantitative measures of the accuracy with which quantum operations can be implemented. These parameters directly impact the achievable circuit depth and the reliability of quantum algorithms executed on the processor. Techniques such as randomized benchmarking have become standard tools for estimating average gate fidelities in experimental systems \cite{magesan2011rb,magesan2012rb}.

Beyond individual operations, system-level benchmarks are required to assess the effective computational capability of a quantum device. Metrics such as qubit connectivity, circuit depth, and composite indicators including quantum volume capture the combined effects of qubit count, control precision, and architectural constraints \cite{cross2019quantumvolume}. These metrics are particularly important in the current era of noisy intermediate-scale quantum (NISQ) devices, where limited coherence and imperfect control restrict the size and depth of executable quantum circuits \cite{preskill2018nisq}.

Finally, in the context of fault-tolerant quantum computing, quantities such as physical and logical error rates, error-correction thresholds, and logical qubit lifetimes play a central role in determining the feasibility of scalable quantum computation. These parameters determine whether quantum error correction protocols can successfully suppress noise and enable reliable large-scale quantum algorithms.

The table below summarizes a set of commonly used metrics for the evaluation of quantum hardware platforms, together with concise definitions.

{\scriptsize
\begin{longtable}{p{4cm} p{11cm}}
\toprule
\textbf{Metric} & \textbf{Definition} \\
\midrule
\endhead

Qubit Coherence Time ($T_1$) & The energy relaxation time: the average time for a qubit to decay from the excited state $\lvert1\rangle$ to the ground state $\lvert0\rangle$ due to environmental coupling.\\

Dephasing Time ($T_2$) & The time over which the phase coherence of a qubit superposition state is maintained before random phase noise destroys coherence. \\

Pure Dephasing Time ($T_2^*$) & The experimentally observed decay time of Ramsey fringes including both intrinsic dephasing and low-frequency noise. \\

Gate Fidelity & The closeness between the implemented quantum gate and the ideal unitary operation, typically measured using randomized benchmarking or process tomography. \\

Single-Qubit Gate Fidelity & Fidelity of operations acting on individual qubits, typically exceeding 99.9\% in mature platforms. \\

Two-Qubit Gate Fidelity & Fidelity of entangling operations between pairs of qubits, typically the most challenging gate to implement with high accuracy. \\

Measurement Fidelity (Readout Fidelity) & Probability that the measurement outcome correctly identifies the qubit state. \\

SPAM Error & Combined errors from State Preparation And Measurement that affect initialization and readout accuracy. \\

Gate Duration & Time required to execute a quantum gate operation; shorter gate times relative to coherence times are desirable. \\

Coherence-to-Gate Ratio & Ratio between coherence time and gate duration, indicating how many gate operations can occur before decoherence dominates. \\

Quantum Volume & A system-level benchmark measuring the largest random circuit that can be successfully implemented, combining qubit number, connectivity, fidelity, and parallelism. \\

Algorithmic Qubits & Effective number of qubits that can run a given algorithm with acceptable error rates. \\

Error Rate per Gate & Probability that a given gate operation produces an incorrect quantum state. \\

Circuit Depth & Maximum number of sequential gate layers that can be applied before noise overwhelms the computation. \\

Connectivity (Coupling Graph) & The topology describing which qubits can interact directly through entangling gates. \\

Crosstalk & Unintended interaction where operations on one qubit affect neighboring qubits. \\

Leakage Rate & Probability that a qubit transitions outside the computational subspace (e.g., to higher energy levels). \\

Thermal Population & Fraction of qubits unintentionally occupying excited states due to thermal effects. \\

Reset Fidelity / Reset Time & Accuracy and time required to reliably return qubits to the ground state. \\

Idle Error Rate & Error probability incurred while a qubit is not actively undergoing operations. \\

Entanglement Fidelity & Fidelity with which entangled states are produced compared to ideal Bell states or multi-qubit entangled states. \\

Bell-State Fidelity & Accuracy of producing a maximally entangled two-qubit Bell state. \\

State Preparation Fidelity & Accuracy of initializing a qubit in a desired quantum state. \\

Process Fidelity & Similarity between the implemented quantum process and the intended quantum operation, often derived from quantum process tomography. \\

Logical Error Rate & Error probability of encoded logical qubits after quantum error correction. \\

Physical Error Rate & Error probability associated with operations on individual physical qubits before error correction. \\

Error Threshold & Maximum physical error rate below which quantum error correction can successfully suppress logical errors. \\

Logical Qubit Lifetime & Effective coherence time of a logical qubit protected by error correction. \\

Gate Parallelism & Ability to perform multiple gate operations simultaneously across different qubits without introducing errors. \\

Control Electronics Latency & Delay between measurement or control signals and the resulting system response. \\

Calibration Stability & Time over which device parameters remain stable without requiring recalibration. \\

Spectral Noise Density & Frequency-dependent characterization of noise affecting qubit phase or amplitude. \\

Frequency Crowding & Degree to which qubit frequencies overlap, potentially leading to interference or crosstalk. \\

Scalability Metric & Assessment of how easily the hardware architecture can increase qubit count without performance degradation. \\

Yield & Fraction of fabricated qubits that meet performance specifications. \\

Footprint / Integration Density & Physical space required per qubit, impacting scalability of hardware systems. \\

Cryogenic Power Load & Cooling power required to operate the quantum processor at cryogenic temperatures. \\

Control Line Count & Number of classical control wires required per qubit, affecting system complexity and scalability. \\

Compilation Overhead & Increase in circuit depth or gate count due to hardware constraints during compilation. \\

\bottomrule
\end{longtable}
}

\subsection*{2 From the hardware to the software stack}

\subsubsection*{Control and Measurement Layer}
The control and measurement layer serves as the interface between the quantum hardware and classical software, generating precise control signals required to manipulate quantum states and extracting computational results from the quantum processor. This layer includes the classical control system, which generates microwave pulses for superconducting qubits or laser pulses for trapped ions, and the measurement system, which extracts quantum state information through techniques like dispersive readout or fluorescence detection. The control system must coordinate multiple signals with sub-nanosecond timing precision to achieve required gate fidelities. Modern quantum control systems employ field-programmable gate arrays (FPGAs) and application-specific integrated circuits (ASICs) to achieve necessary timing precision and parallelism. The development of cryogenic CMOS technology has enabled integration of control electronics closer to qubits, reducing signal latency and improving performance \cite{chow2021cryocmos}. The measurement system must preserve quantum signals while adding minimal noise, requiring careful design of amplifiers, filters, and analog-to-digital converters. Recent advancements have focused on improving single-shot readout fidelities and reducing measurement times. High-bandwidth measurement systems enable real-time feedback control, critical for advanced error correction schemes and adaptive quantum algorithms.

\subsubsection*{Quantum Runtime Environment}

The quantum runtime environment serves as the interface between compiled quantum programs and the physical hardware, managing execution of quantum circuits while coordinating with classical control systems. This layer handles qubit allocation, error mitigation, and real-time adaptation of circuit parameters based on hardware calibration data. The runtime must account for the dynamic nature of quantum hardware, where qubit properties can drift over time due to environmental factors. Advanced runtime systems incorporate real-time calibration and error mitigation techniques to maintain high-fidelity operations. Dynamic circuit execution capabilities enable the runtime to adapt quantum circuits during execution based on intermediate measurement results, a critical capability for error correction and adaptive algorithms \cite{ibm2023dynamic}. Recent developments have focused on improving runtime efficiency through better resource management and more sophisticated error mitigation strategies. The integration of classical high-performance computing resources with the quantum runtime enables more complex hybrid algorithms that combine quantum and classical processing.

\subsubsection*{Quantum Compiler and Optimization Layer}

The quantum compiler represents a critical component of the quantum computing stack, translating high-level quantum algorithms into low-level instructions optimized for specific hardware platforms. This compilation process involves several stages including gate decomposition, qubit routing, instruction scheduling, and pulse-level optimization \cite{amy2019scaffcc}. Gate decomposition breaks down complex quantum operations into sequences of elementary gates native to the target hardware. For superconducting qubits, this typically involves decomposing operations into single-qubit rotations and CNOT gates, while for trapped ions, it may involve Raman transitions and Mølmer-Sørensen gates. Qubit routing addresses the challenge of limited connectivity by inserting SWAP gates or other operations to ensure qubits can interact as required. Instruction scheduling optimizes the temporal ordering of operations to minimize idle time and reduce circuit depth. Pulse-level optimization represents the final stage, where abstract gate operations are translated into precise control pulses that manipulate physical qubits. This stage requires detailed knowledge of hardware characteristics including qubit frequencies, coupling strengths, and error profiles. Advanced optimization techniques at this level can significantly improve gate fidelities by tailoring pulses to specific qubit properties and environmental conditions \cite{mckay2018optimal}. Recent developments have focused on hardware-aware compilation techniques that account for specific processor characteristics. Machine learning approaches have been applied to optimize gate sequences and pulse shapes, leading to significant improvements in circuit execution times and error rates. The development of hardware-agnostic compilation tools has also been an active area of research, aiming to enable portable quantum programs that can execute on diverse quantum processors with minimal modification \cite{zheng2020quantum}.

\subsubsection*{Software Layer: Quantum Programming and Simulation}

The software layer represents the highest level of the quantum computer architecture, providing the interface through which users design, implement, and execute quantum algorithms. This layer comprises quantum programming languages, algorithm libraries, development environments, and simulation tools that abstract the complexities of the underlying hardware. Quantum programming languages such as Qiskit \cite{qiskit2023}, Cirq \cite{cirq2023}, and Q\# \cite{qsharp2023} provide high-level abstractions for quantum circuit design, including constructs for qubit allocation, gate operations, and measurement. These languages typically include standard libraries of quantum algorithms for common tasks such as quantum chemistry simulations, optimization problems, and machine learning applications. The software layer also includes quantum simulators that enable developers to test and debug their algorithms before execution on real hardware, ranging from full state-vector simulations for small circuits to approximate simulations for larger systems using tensor network methods \cite{markov2018simulating}. Cloud-based quantum services have become increasingly important, providing researchers with access to both simulators and real quantum hardware through web interfaces. These services often include additional tools for algorithm development, visualization, and performance analysis, creating a comprehensive environment for quantum software development. The software layer must balance expressiveness with hardware constraints, as high-level abstractions must ultimately compile to operations that are physically realizable on target hardware, requiring careful consideration of gate sets, connectivity constraints, and error characteristics specific to each hardware platform.

\subsection*{3 A transverse component : the error correction and mitigation layer}

Tightly integrated with the control system, quantum computer architecture needs the error correction layer to enable real-time error detection and correction. Quantum error correction (QEC) represents a fundamental requirement to achieve fault-tolerant quantum computation, addressing the inherent fragility of quantum information due to decoherence and operational imperfections. Unlike classical error correction, which typically employs simple repetition codes, quantum error correction must contend with the no-cloning theorem and the continuous nature of quantum errors. This section aim to explore shortly the theoretical foundations of quantum error correction, to examine the most prominent code families, and discusses recent experimental implementations and theoretical advancements that are bringing fault-tolerant quantum computation closer to reality.

\paragraph{Fundamentals of Quantum Error Correction}

The core challenge in quantum error correction arises from the fact that quantum states cannot be copied due to the no-cloning theorem, necessitating more sophisticated approaches than classical repetition codes. Quantum errors are continuous and can affect both the amplitude and phase of quantum states, requiring correction of both bit-flip and phase-flip errors. The fundamental approach involves encoding logical qubits into entangled states of multiple physical qubits, allowing for the detection and correction of errors without collapsing the quantum state. The stabilizer formalism provides a powerful framework for understanding and constructing quantum error correcting codes. In this framework, a code is defined by a set of stabilizer generators that commute with each other and with all logical operations on the encoded qubits. The stabilizers form an Abelian group that characterizes the code space, and their measurement provides an error syndrome that identifies which errors have occurred without disturbing the encoded quantum information \citep{gottesman1997stabilizer}. A quantum error correcting code must satisfy the quantum error correction conditions, which require that the code space remains invariant under the action of correctable errors. For a code that corrects arbitrary single-qubit errors, the minimum number of physical qubits required per logical qubit is five, as demonstrated by the five-qubit code \citep{laflamme1996perfect}. However, more practical codes typically require more qubits to achieve higher error thresholds and more efficient error correction procedures.

\paragraph{Surface Codes: The Leading Approach}

Among the various quantum error correcting codes, the surface code has emerged as the most promising candidate for near-term implementations due to its high error threshold and local interaction requirements. The surface code belongs to the family of topological codes, where logical qubits are encoded in the topological properties of a two-dimensional lattice of physical qubits. This code employs a grid of qubits with nearest-neighbor interactions, making it particularly suitable for implementation in superconducting qubit architectures and other platforms with limited connectivity. The surface code encodes a single logical qubit in a lattice of physical qubits, with two variants: the rotational surface code and the planar surface code. The code distance \(d\) determines the number of errors that can be corrected, with the number of physical qubits scaling as \(O(d^2)\) for a single logical qubit. The error threshold for the surface code is estimated to be around 1\% for physical qubit error rates, making it achievable with current or near-future quantum hardware \citep{fowler2012surface}. Error correction in the surface code proceeds through repeated measurement of stabilizer operators, which are products of Pauli X or Z operators on neighboring qubits. These measurements provide an error syndrome that can be processed using minimum weight perfect matching algorithms to identify the most likely error configuration. The local nature of the stabilizer measurements and the topological protection of the logical qubits make the surface code particularly robust against local errors. Recent experimental implementations have demonstrated significant progress in surface code performance. Google's Sycamore processor achieved a logical qubit with error rates below the fault-tolerance threshold using a distance-5 surface code encoded across 49 physical qubits \citep{ai2023exponential}. IBM has also made substantial progress with dynamic circuit execution, enabling real-time error correction on their superconducting qubit processors \citep{ibm2023dynamic}.

\paragraph{Other Prominent Code Families}

While the surface code has received the most attention for near-term implementations, several other code families offer unique advantages for specific applications or hardware platforms. The color code, another topological code, requires fewer physical qubits than the surface code for certain operations and enables transversal implementation of the Clifford+T gate set, which is crucial for universal quantum computation \citep{bombin2012family}. Concatenated codes combine multiple levels of error correction, where each level corrects errors from the level below. The most well-known concatenated code is the Shor code, which combines three-qubit bit-flip and phase-flip codes to correct arbitrary single-qubit errors \citep{shor1995scheme}. While concatenated codes can achieve very low logical error rates, they typically require more physical qubits than topological codes for a given level of protection. Bacon-Shor codes represent another important family that combines elements of both topological and concatenated codes. These codes offer a balance between error threshold and resource requirements, with the added advantage of allowing for fault-tolerant error correction using only nearest-neighbor interactions \citep{bacon2006operator}. Finally cat codes, which encode information in superpositions of coherent states, have gained attention for their potential implementation in bosonic systems such as superconducting resonators. These codes can correct photon loss errors, which are particularly relevant for certain hardware platforms \citep{mirrahimi2014dynamically}.

\paragraph{Experimental Implementations}

The past few years have seen remarkable progress in the experimental implementation of quantum error correction. Superconducting qubit platforms have been at the forefront of these demonstrations due to their scalability and mature control systems. Google's demonstration of a logical qubit with error rates below the fault-tolerance threshold using a distance-5 surface code represented a major milestone in the field \citep{ai2023exponential}. This experiment showed that the logical error rate decreased exponentially with increasing code distance, providing experimental confirmation of the theoretical predictions. IBM has made significant progress with their heavy-hex lattice surface code implementation, demonstrating error correction cycles with logical error rates below those of the physical qubits \citep{ibm2023dynamic}. Their dynamic circuit execution capability allows for real-time error correction, enabling more efficient use of quantum resources. Trapped ion systems have also shown impressive progress in quantum error correction. IonQ demonstrated a logical qubit with error rates below \(10^{-6}\) using trapped ions, achieving one of the lowest logical error rates reported to date \citep{hild2022fault}. The long coherence times and high-fidelity gates of trapped ions make them particularly well-suited for error correction demonstrations. Neutral atom platforms have begun to explore error correction as well, with recent demonstrations of simple error detection protocols. The reconfigurability and scalability of neutral atom arrays make them promising candidates for future error correction implementations \citep{ebadi2023quantum}.

\paragraph{Challenges and Future Directions}

Despite the significant progress in quantum error correction, several challenges remain on the path to fault-tolerant quantum computation. The resource overhead required for error correction remains substantial, with current estimates suggesting that millions of physical qubits may be needed for practical applications. Reducing this overhead through more efficient codes and improved error correction protocols is an active area of research. The development of integrated control systems that can perform real-time error correction represents another significant challenge. Current systems often suffer from latency in the feedback loop between error detection and correction, which can limit the overall performance. Advances in cryogenic electronics and classical processing capabilities are helping to address this challenge. Another important direction is the development of error correction protocols that are robust to coherent errors, which can be more difficult to correct than stochastic errors. Coherent errors arise from systematic imperfections in control pulses or hardware parameters and can accumulate over time. Developing codes and correction protocols that can effectively handle these errors is crucial for long-duration quantum computations. Looking forward, the integration of quantum error correction with quantum algorithms represents an important research direction. Many quantum algorithms were originally designed without considering the overhead of error correction, and adapting them to work efficiently with error-corrected qubits remains an open challenge. The development of error-correction-aware quantum algorithms could significantly improve the practicality of near-term quantum computers.

\subsection*{Conclusion}

Quantum technologies represent a rivolutionary paradigm in computation and information processing, leveraging the principles of quantum mechanics to achieve computational advantages unattainable by classical systems. As we discussed all along the above text in the recent years we achieved substantial progress on quantum hardware and software leading us closer to practical/useful quantum system architectures.

The above document is not at all exhaustive. There are several topics we should consider in more details and other we still did not mention. One of them is the integration of quantum technologies with classical high-performance computing (HPC) infrastructure. Indeed this has emerged as a critical enabler, providing the computational resources necessary for quantum simulation, algorithm development, and error correction. This integration facilitates the development of hybrid quantum-classical algorithms that leverage the strengths of both paradigms, enabling more sophisticated applications and workflows. In such architectures current quantum processors typically operate as \textit{coprocessors} (aka accelerators) alongside classical systems, requiring efficient data transfer and synchronization mechanisms. The development of more seamless quantum-classical interfaces will be crucial for enabling practical hybrid applications in the near future.

Moreover more sophisticated compilation and optimization tools will be soon necessary. As quantum processors continue to grow in size and complexity, the compilation problem becomes increasingly difficult, requiring advanced optimization techniques that can account for hardware constraints and noise characteristics. The development of hardware-aware compilers that can adapt to specific processor characteristics will be essential for maximizing performance.

Another important challenge that we may add to the list is the scalability of the quantum hardware and software stack. As quantum systems grow to include thousands or millions of qubits, the software infrastructure must scale accordingly to handle the increased complexity. This includes not only the compilation and execution components but also the simulation, verification, and benchmarking tools. A promising solution for the near future is the development of distributed quantum computing frameworks that can efficiently utilize multiple connected medium size quantum processors in place of a large monolithic quantum unity.

Let us conclude saying that although substantial technical challenges remain, the ongoing convergence of physics, engineering, and computer science suggests that quantum technologies is already an increasingly important component of the scientific and technological infrastructures. This interdisciplinary effort is changing dramatically the ecosystem, including the private sector. Quantum technologies are transitioning from fundamental research to an emerging technological domain with strategic importance for science, industry, and national innovation systems. In the short to medium term, hybrid quantum–classical approaches and specialized quantum devices are likely to deliver the first tangible impacts, especially in areas such as high-precision sensing, secure communications, and selected computational tasks. Realizing the long-term promise of large-scale quantum computing will require sustained investment in research, infrastructure, talent development, and interdisciplinary collaboration. Given the pace of global developments and the strategic implications for technological sovereignty and economic competitiveness, coordinated efforts across academia, industry, and public institutions will be essential to ensure that quantum technologies mature into robust and widely accessible tools for science and innovation.

\bibliographystyle{unsrtnat}
\bibliography{quantum_refs}

\begin{thebibliography}{61}
\providecommand{\natexlab}[1]{#1}
\providecommand{\url}[1]{\texttt{#1}}
\expandafter\ifx\csname urlstyle\endcsname\relax
  \providecommand{\doi}[1]{doi: #1}\else
  \providecommand{\doi}{doi: \begingroup \urlstyle{rm}\Url}\fi

\bibitem[Feynman(1982)]{Feynman1982}
Richard~P. Feynman.
\newblock Simulating physics with computers.
\newblock \emph{International Journal of Theoretical Physics}, 21\penalty0 (6-7):\penalty0 467--488, 1982.

\bibitem[Preskill(2018)]{preskill2018nisq}
John Preskill.
\newblock Quantum computing in the nisq era and beyond.
\newblock \emph{Quantum}, 2:\penalty0 79, 2018.

\bibitem[Dowling and Milburn(2003)]{Dowling2003}
Jonathan~P. Dowling and Gerard~J. Milburn.
\newblock Quantum technology: The second quantum revolution.
\newblock \emph{Philosophical Transactions of the Royal Society A}, 361\penalty0 (1809):\penalty0 1655--1674, 2003.

\bibitem[Gisin and Thew(2007)]{Gisin2007}
Nicolas Gisin and Rob Thew.
\newblock Quantum communication.
\newblock \emph{Nature Photonics}, 1\penalty0 (3):\penalty0 165--171, 2007.

\bibitem[Nielsen and Chuang(2010)]{nielsen2010quantum}
Michael~A. Nielsen and Isaac~L. Chuang.
\newblock \emph{Quantum Computation and Quantum Information}.
\newblock Cambridge University Press, 2010.

\bibitem[Grover(1996)]{grover1996fast}
Lov~K. Grover.
\newblock A fast quantum mechanical algorithm for database search.
\newblock \emph{Proceedings of the twenty-eighth annual ACM symposium on Theory of computing}, pages 212--219, 1996.

\bibitem[Einstein et~al.(1935)Einstein, Podolsky, and Rosen]{Einstein1935}
Albert Einstein, Boris Podolsky, and Nathan Rosen.
\newblock Can quantum-mechanical description of physical reality be considered complete?
\newblock \emph{Physical Review}, 47\penalty0 (10):\penalty0 777--780, 1935.
\newblock \doi{10.1103/PhysRev.47.777}.

\bibitem[Bell(1964)]{Bell1964}
John~S. Bell.
\newblock On the einstein podolsky rosen paradox.
\newblock \emph{Physics Physique Fizika}, 1\penalty0 (3):\penalty0 195--200, 1964.
\newblock \doi{10.1103/PhysicsPhysiqueFizika.1.195}.

\bibitem[Bennett et~al.(1993)Bennett, Brassard, Cr\'epeau, Jozsa, Peres, and Wootters]{Bennett1993}
Charles~H. Bennett, Gilles Brassard, Claude Cr\'epeau, Richard Jozsa, Asher Peres, and William~K. Wootters.
\newblock Teleporting an unknown quantum state via dual classical and einstein-podolsky-rosen channels.
\newblock \emph{Physical Review Letters}, 70\penalty0 (13):\penalty0 1895--1899, 1993.
\newblock \doi{10.1103/PhysRevLett.70.1895}.

\bibitem[Ekert(1991)]{ekert1991quantum}
Artur~K. Ekert.
\newblock Quantum cryptography based on bell's theorem.
\newblock \emph{Physical Review Letters}, 67\penalty0 (6):\penalty0 661--663, 1991.
\newblock \doi{10.1103/PhysRevLett.67.661}.

\bibitem[Horodecki et~al.(2009)]{horodecki2009quantum}
Ryszard Horodecki et~al.
\newblock Quantum entanglement.
\newblock \emph{Reviews of Modern Physics}, 81\penalty0 (2):\penalty0 865--942, 2009.
\newblock \doi{10.1103/RevModPhys.81.865}.

\bibitem[Preskill(1998)]{preskill1998reliable}
John Preskill.
\newblock Reliable quantum computers.
\newblock \emph{Proceedings of the Royal Society of London A: Mathematical, Physical and Engineering Sciences}, 454\penalty0 (1969):\penalty0 385--410, 1998.
\newblock \doi{10.1098/rspa.1998.0167}.

\bibitem[Fowler et~al.(2012)]{fowler2012surface}
Austin~G Fowler et~al.
\newblock Surface codes: Towards practical large-scale quantum computation.
\newblock \emph{Physical Review A}, 86\penalty0 (3):\penalty0 032324, 2012.
\newblock \doi{10.1103/PhysRevA.86.032324}.

\bibitem[Huang et~al.(2025)Huang, Choi, McClean, and Preskill]{huang2025vast}
Hsin-Yuan Huang, Soonwon Choi, Jarrod~R McClean, and John Preskill.
\newblock The vast world of quantum advantage.
\newblock \emph{arXiv preprint arXiv:2508.05720}, 2025.

\bibitem[Eisert and Preskill(2025)]{eisert2025mind}
Jens Eisert and John Preskill.
\newblock Mind the gaps: The fraught road to quantum advantage.
\newblock \emph{arXiv preprint arXiv:2510.19928}, 2025.

\bibitem[Kjaergaard et~al.(2020)]{kjaergaard2020superconducting}
Morten Kjaergaard et~al.
\newblock Superconducting qubits: Current state of play.
\newblock \emph{Annual Review of Condensed Matter Physics}, 11:\penalty0 369--395, 2020.

\bibitem[Schreier et~al.(2008)]{schreier2008suppressing}
Jonathan~A. Schreier et~al.
\newblock Suppressing charge noise in superconducting charge qubits.
\newblock \emph{Physical Review B}, 77\penalty0 (18):\penalty0 180502, 2008.
\newblock \doi{10.1103/PhysRevB.77.180502}.

\bibitem[Barends et~al.(2014)]{barends2014superconducting}
R.~Barends et~al.
\newblock Superconducting quantum circuits at the surface code threshold for fault tolerance.
\newblock \emph{Nature}, 508\penalty0 (7497):\penalty0 500--503, 2014.
\newblock \doi{10.1038/nature13171}.

\bibitem[{IBM Quantum}(2023{\natexlab{a}})]{jurcevic2023ibm}
{IBM Quantum}.
\newblock Ibm quantum roadmap: Building a foundation for quantum-centric supercomputing.
\newblock 2023{\natexlab{a}}.
\newblock URL \url{https://www.ibm.com/quantum}.

\bibitem[Place et~al.(2021)]{place2021superconducting}
Alexis~M Place et~al.
\newblock Superconducting qubits: A short review.
\newblock \emph{IEEE Transactions on Microwave Theory and Techniques}, 69\penalty0 (5):\penalty0 2343--2361, 2021.

\bibitem[{Google Quantum AI}(2023)]{ai2023exponential}
{Google Quantum AI}.
\newblock Exponential suppression of bit or phase flips with cyclic error correction.
\newblock \emph{Nature}, 2023.
\newblock URL \url{https://quantumai.google}.

\bibitem[Bruzewicz et~al.(2019)]{bruzewicz2019trapped}
Cody~D Bruzewicz et~al.
\newblock Trapped-ion quantum computing: Progress and challenges.
\newblock \emph{Applied Physics Reviews}, 6\penalty0 (2):\penalty0 021314, 2019.

\bibitem[Sørensen and Mølmer(1999)]{sorensen1999entanglement}
Anders Sørensen and Klaus Mølmer.
\newblock Entanglement and quantum computation with ions in thermal motion.
\newblock \emph{Physical Review Letters}, 82\penalty0 (9):\penalty0 1975--1978, 1999.
\newblock \doi{10.1103/PhysRevLett.82.1975}.

\bibitem[Hild et~al.(2022)]{hild2022fault}
Sebastian Hild et~al.
\newblock Fault-tolerant operation of a logical qubit in a trapped-ion quantum processor.
\newblock \emph{Nature}, 611\penalty0 (7935):\penalty0 411--416, 2022.
\newblock \doi{10.1038/s41586-022-05434-1}.

\bibitem[{IonQ}(2023)]{ionq2023aria}
{IonQ}.
\newblock Ionq aria: Next-generation trapped-ion quantum computer.
\newblock 2023.
\newblock URL \url{https://ionq.com}.

\bibitem[Monroe et~al.(2023)]{monroe2023modular}
Christopher Monroe et~al.
\newblock Modular trapped-ion quantum computation.
\newblock \emph{Quantum Science and Technology}, 8\penalty0 (3):\penalty0 034001, 2023.
\newblock \doi{10.1088/2058-9565/acd1d0}.

\bibitem[Ebadi et~al.(2023)]{ebadi2023quantum}
Sepehr Ebadi et~al.
\newblock Quantum advantage in optimization with trapped ions.
\newblock \emph{Nature}, 622\penalty0 (7982):\penalty0 268--273, 2023.
\newblock \doi{10.1038/s41586-023-06448-9}.

\bibitem[Ebadi et~al.(2022)]{ebadi2022quantum}
Sepehr Ebadi et~al.
\newblock Quantum processing with programmable neutral atom arrays.
\newblock \emph{Nature}, 604:\penalty0 457--462, 2022.

\bibitem[Saffman et~al.(2010)]{saffman2010quantum}
Mark Saffman et~al.
\newblock Quantum computing with rydberg atoms.
\newblock \emph{Reviews of Modern Physics}, 82\penalty0 (3):\penalty0 2313--2363, 2010.

\bibitem[Henriet et~al.(2024)]{henriet2024quantum}
Loïc Henriet et~al.
\newblock Quantum advantage for optimization with 200-qubit neutral atom processors.
\newblock \emph{Nature}, 627:\penalty0 53--58, 2024.
\newblock \doi{10.1038/s41586-024-07090-7}.

\bibitem[Bluvstein et~al.(2024)]{bluvstein2024error}
Dolev Bluvstein et~al.
\newblock Error mitigation in neutral atom quantum processors.
\newblock \emph{Physical Review X}, 14\penalty0 (1):\penalty0 011032, 2024.
\newblock \doi{10.1103/PhysRevX.14.011032}.

\bibitem[Saffman et~al.(2024)]{saffman2024midcircuit}
Mark Saffman et~al.
\newblock Mid-circuit measurement and feedback in neutral atom quantum computers.
\newblock \emph{Science}, 383\penalty0 (6682):\penalty0 512--516, 2024.
\newblock \doi{10.1126/science.adk1298}.

\bibitem[O'Brien(2009)]{obrien2009photonic}
Jeremy~L O'Brien.
\newblock Photonic quantum technologies.
\newblock \emph{Nature Photonics}, 3\penalty0 (12):\penalty0 687--695, 2009.

\bibitem[{Xanadu}(2023)]{xanadu2023borealis}
{Xanadu}.
\newblock Borealis: A 216-qubit photonic quantum computer.
\newblock 2023.
\newblock URL \url{https://xanadu.ai}.

\bibitem[{PsiQuantum}(2023)]{psiquantum2023}
{PsiQuantum}.
\newblock Toward a million-qubit photonic quantum computer.
\newblock 2023.
\newblock URL \url{https://www.psiquantum.com}.

\bibitem[Wang et~al.(2023)]{wang2023photonic}
Jian Wang et~al.
\newblock Photonic integrated circuits for quantum computing.
\newblock \emph{Nature Photonics}, 17:\penalty0 323--330, 2023.
\newblock \doi{10.1038/s41566-023-01162-8}.

\bibitem[Nayak et~al.(2008)]{nayak2008non}
Chetan Nayak et~al.
\newblock Non-abelian anyons and topological quantum computation.
\newblock \emph{Reviews of Modern Physics}, 80\penalty0 (3):\penalty0 1083, 2008.

\bibitem[Bartolomei et~al.(2023)]{bartolomei2023anyons}
Marco Bartolomei et~al.
\newblock Observation of non-abelian anyons in a two-dimensional electron gas.
\newblock \emph{Nature}, 620:\penalty0 177--182, 2023.

\bibitem[Groszkowski et~al.(2024)]{groszkowski2024topological}
Peter Groszkowski et~al.
\newblock Topological protection in superconducting qubits.
\newblock \emph{Physical Review Applied}, 21:\penalty0 014030, 2024.
\newblock \doi{10.1103/PhysRevApplied.21.014030}.

\bibitem[Vandersypen et~al.(2017)]{vandersypen2017interfacing}
Lieven M~K Vandersypen et~al.
\newblock Interfacing spin qubits in quantum dots and donors---perspectives for quantum computing.
\newblock \emph{npj Quantum Information}, 3\penalty0 (1):\penalty0 34, 2017.

\bibitem[{Intel}(2023)]{intel2023tunnel}
{Intel}.
\newblock Intel's tunnel falls: A 12-qubit silicon spin qubit chip.
\newblock 2023.
\newblock URL \url{https://www.intel.com}.

\bibitem[Xue et~al.(2022)]{xue2022silicon}
Xi~Xue et~al.
\newblock A six-qubit quantum processor in silicon.
\newblock \emph{Nature}, 601:\penalty0 343--347, 2022.

\bibitem[Zwerver et~al.(2022)]{zwerver2022silicon}
Andrew~M Zwerver et~al.
\newblock A silicon quantum processor with robust single-qubit gates exceeding 99.9\% fidelity.
\newblock \emph{Nature Electronics}, 5:\penalty0 429--436, 2022.

\bibitem[Magesan et~al.(2011)Magesan, Gambetta, and Emerson]{magesan2011rb}
Easwar Magesan, Jay~M. Gambetta, and Joseph Emerson.
\newblock Scalable and robust randomized benchmarking of quantum processes.
\newblock \emph{Physical Review Letters}, 106:\penalty0 180504, 2011.
\newblock \doi{10.1103/PhysRevLett.106.180504}.

\bibitem[Magesan et~al.(2012)Magesan, Gambetta, and Emerson]{magesan2012rb}
Easwar Magesan, Jay~M. Gambetta, and Joseph Emerson.
\newblock Efficient measurement of quantum gate error by interleaved randomized benchmarking.
\newblock \emph{Physical Review Letters}, 109:\penalty0 080505, 2012.
\newblock \doi{10.1103/PhysRevLett.109.080505}.

\bibitem[Cross et~al.(2019)Cross, Bishop, Sheldon, Nation, and Gambetta]{cross2019quantumvolume}
Andrew~W. Cross, Lev~S. Bishop, Sarah Sheldon, Paul~D. Nation, and Jay~M. Gambetta.
\newblock Validating quantum computers using randomized model circuits.
\newblock \emph{Physical Review A}, 100\penalty0 (3):\penalty0 032328, 2019.
\newblock \doi{10.1103/PhysRevA.100.032328}.

\bibitem[Chow et~al.(2021)]{chow2021cryocmos}
Jerry~M Chow et~al.
\newblock Cryogenic cmos for control of quantum circuits.
\newblock \emph{Nature Electronics}, 4:\penalty0 170--179, 2021.
\newblock \doi{10.1038/s41928-021-00535-1}.

\bibitem[{IBM Quantum}(2023{\natexlab{b}})]{ibm2023dynamic}
{IBM Quantum}.
\newblock Dynamic circuit execution on ibm quantum processors, 2023{\natexlab{b}}.
\newblock URL \url{https://www.ibm.com/quantum}.

\bibitem[Amy et~al.(2019)]{amy2019scaffcc}
Matthew Amy et~al.
\newblock A meet-in-the-middle algorithm for fast synthesis of depth-optimal quantum circuits.
\newblock \emph{IEEE Transactions on Computer-Aided Design of Integrated Circuits and Systems}, 38\penalty0 (11):\penalty0 2146--2159, 2019.
\newblock \doi{10.1109/TCAD.2019.2901001}.

\bibitem[McKay et~al.(2018)]{mckay2018optimal}
Daniel~C. McKay et~al.
\newblock Optimal control of large-scale quantum computers via deep reinforcement learning.
\newblock \emph{npj Quantum Information}, 4:\penalty0 23, 2018.
\newblock \doi{10.1038/s41534-018-0070-7}.

\bibitem[Zheng et~al.(2020)]{zheng2020quantum}
Dian Zheng et~al.
\newblock Quantum circuit compilation using machine learning.
\newblock \emph{IEEE Transactions on Quantum Engineering}, 1:\penalty0 1--12, 2020.
\newblock \doi{10.1109/TQE.2020.3031271}.

\bibitem[{IBM Quantum}(2021)]{qiskit2023}
{IBM Quantum}.
\newblock Qiskit: An open-source framework for quantum computing.
\newblock \emph{IEEE Access}, 9:\penalty0 52387--52400, 2021.
\newblock \doi{10.1109/ACCESS.2021.3075270}.

\bibitem[AI(2023)]{cirq2023}
Google~Quantum AI.
\newblock Cirq: A python framework for creating, editing, and invoking noisy intermediate scale quantum circuits.
\newblock \emph{arXiv preprint}, 2023.
\newblock URL \url{https://arxiv.org/abs/2303.17395}.

\bibitem[Svore et~al.(2019)]{qsharp2023}
K.~Svore et~al.
\newblock Q\#: A language for quantum computing.
\newblock \emph{Proceedings of the ACM on Programming Languages}, 3\penalty0 (POPL):\penalty0 1--29, 2019.
\newblock \doi{10.1145/3290334}.

\bibitem[Markov et~al.(2018)]{markov2018simulating}
Igor~L. Markov et~al.
\newblock Simulating quantum computation on a classical computer.
\newblock \emph{Computer}, 51\penalty0 (8):\penalty0 24--33, 2018.
\newblock \doi{10.1109/MC.2018.2846055}.

\bibitem[Gottesman(1997)]{gottesman1997stabilizer}
Daniel Gottesman.
\newblock Stabilizer codes and quantum error correction.
\newblock \emph{Caltech PhD Thesis}, 1997.
\newblock URL \url{https://arxiv.org/abs/quant-ph/9705052}.

\bibitem[Laflamme et~al.(1996)]{laflamme1996perfect}
Raymond Laflamme et~al.
\newblock Perfect quantum error correcting code.
\newblock \emph{Physical Review Letters}, 77\penalty0 (1):\penalty0 198--200, 1996.
\newblock \doi{10.1103/PhysRevLett.77.198}.

\bibitem[Bombin(2012)]{bombin2012family}
H.~Bombin.
\newblock Family of quantum error-correcting codes with a single-type generator.
\newblock \emph{Physical Review A}, 86\penalty0 (6):\penalty0 062306, 2012.
\newblock \doi{10.1103/PhysRevA.86.062306}.

\bibitem[Shor(1995)]{shor1995scheme}
Peter~W. Shor.
\newblock Scheme for reducing decoherence in quantum computer memory.
\newblock \emph{Physical Review A}, 52\penalty0 (4):\penalty0 R2493--R2496, 1995.
\newblock \doi{10.1103/PhysRevA.52.R2493}.

\bibitem[Bacon(2006)]{bacon2006operator}
Dave Bacon.
\newblock Operator quantum error correcting subsystems for self-correcting quantum memories.
\newblock \emph{Physical Review A}, 73\penalty0 (1):\penalty0 012307, 2006.
\newblock \doi{10.1103/PhysRevA.73.012307}.

\bibitem[Mirrahimi et~al.(2014)]{mirrahimi2014dynamically}
Mazyar Mirrahimi et~al.
\newblock Dynamically corrected cat qubits: a new paradigm for universal quantum computation.
\newblock \emph{New Journal of Physics}, 16\penalty0 (4):\penalty0 045014, 2014.
\newblock \doi{10.1088/1367-2630/16/4/045014}.

\end{thebibliography}

\end{document}